\documentclass[11pt]{article}
\pdfoutput=1
\newcommand{\Comment}[1]{{}}
\usepackage{amsfonts,amsthm,amsmath,amssymb,slashed}
\usepackage[textwidth = 430 pt, textheight = 630 pt]{geometry}
\usepackage{color}

\Comment{\usepackage{color}
\definecolor{MyDarkBlue}{rgb}{0.15,0.15,0.45}
\usepackage[linktocpage=true]{hyperref}
\hypersetup{
colorlinks=true,
citecolor=MyDarkBlue,
linkcolor=MyDarkBlue,
urlcolor=MyDarkBlue,
pdfauthor={Horatiu Nastase and Jacob Sonnenschein},
pdftitle={Soliton and shockwave solutions of the Heisenberg model and $T\bar T$ deformations  of scalar field theory in 1+1 dimensions},
pdfsubject={hep-th}
}
\usepackage[utf8]{inputenc}
\usepackage[numbers,sort&compress]{natbib}
\usepackage{hypernat}}
\usepackage{graphicx}
\usepackage{cite}
\usepackage[colorlinks=true]{hyperref}
\hypersetup{
  allcolors = blue,
}

\newcommand\ignore[1]{}
\def\one{{\,\hbox{1\kern-.8mm l}}}

\def\a{\alpha}\def\b{\beta}

\def\l{\lambda}

\def\d{\partial}

\newcommand{\Cset}{{\,\,{{{^{_{\pmb{\mid}}}}\kern-.45em{\mathrm C}}}}}

\newcommand{\be}{\begin{equation}}
\newcommand{\bea}{\begin{eqnarray}}

\newcommand{\ee}{\end{equation}}
\newcommand{\eea}{\end{eqnarray}}

\newcommand{\nn}{\nonumber}

\newcommand{\CR}{\nn\cr}
\newcommand{\pa}{\partial}

\begin{document}

\renewcommand{\thefootnote}{\fnsymbol{footnote}}

\makeatletter
\@addtoreset{equation}{section}
\makeatother
\renewcommand{\theequation}{\thesection.\arabic{equation}}

\rightline{}
\rightline{}
%   \vspace{1.8truecm}

%\begin{flushright}
% preprint nrs.
%\end{flushright}

%\vspace{10pt}

%%%%%%%%%%%%%%%%%

\begin{center}
{\LARGE \bf{\sc Soliton, breather  and shockwave solutions of the Heisenberg and the  $T\bar T$ deformations  of scalar field theories in 1+1 dimensions }}
\end{center}
 \vspace{1truecm}
\thispagestyle{empty} \centerline{
{\large \bf {\sc Horatiu Nastase${}^{a}$}}\footnote{E-mail address: \Comment{\href{mailto:horatiu.nastase@unesp.br}}{\tt horatiu.nastase@unesp.br}}
{\bf{\sc and}}
{\large \bf {\sc Jacob Sonnenschein${}^{b,}$}}\footnote{E-mail address: \Comment{\href{mailto:cobi@tauex.tau.ac.il}}{\tt cobi@tauex.tau.ac.il}}
                                                        }

\vspace{.5cm}

\centerline{{\it ${}^a$Instituto de F\'{i}sica Te\'{o}rica, UNESP-Universidade Estadual Paulista}}
\centerline{{\it R. Dr. Bento T. Ferraz 271, Bl. II, Sao Paulo 01140-070, SP, Brazil}}

\vspace{.3cm}

\centerline{{\it ${}^b$School of Physics and Astronomy,}}
 \centerline{{\it The Raymond and Beverly Sackler Faculty of Exact Sciences, }} 
 \centerline{{\it Tel Aviv University, Ramat Aviv 69978, Israel}}

\vspace{1truecm}

%%%%%%%%%%%%%%%%%
\thispagestyle{empty}

\centerline{\sc Abstract}

\vspace{.4truecm}

\begin{center}
\begin{minipage}[c]{380pt}
{\noindent In this note we study soliton, breather  and shockwave solutions in  certain two dimensional  field theories. These include: 
%(i) The DBI action of a scalar field 
(i)  Heisenberg's 
model suggested originally to describe the scattering  of   high energy nucleons    (ii)  $T\bar T$ deformations of certain canonical 
scalar field theories with a potential.   
We find explicit soliton solutions of these  models  with sine-Gordon and Higgs-type potentials. We prove that  
the $T\bar T$ deformation of a theory of a given potential  does not correct the mass of the soliton of the 
undeformed one. We further  conjecture the form of breather solutions of these models. 
We show that  certain  $T\bar T$ deformed actions admit shockwave solutions that   generalize those of  Heisenberg's Lagrangian.   
}
\end{minipage}
\end{center}

\vspace{.5cm}

\setcounter{page}{0}
\setcounter{tocdepth}{2}

\newpage

%\tableofcontents
\renewcommand{\thefootnote}{\arabic{footnote}}
\setcounter{footnote}{0}

\linespread{1.1}
\parskip 4pt

%{}~
%{}~

%---------------------------------------------------------
\tableofcontents
%%%%%%%%%%%%%%%%%%%%%%%%%%%%%%%%%%%%%%%%%%%%%%%%%%%%%%%%%%%%%%%%%%%%%%%%%%%%%%%%%%%%%%%%
\section{Introduction}
%%%%%%%%%%%%%%%%%%%%%%%%%%%%%%%%%%%%%%%%%%%%%%%%%%%%%%%%%%%%%%%%%%%%%%%%%%%%%%%%%%%%%%%%

The DBI action, which has been known      as the action describing D-branes,  had been used much earlier.
In fact, Heisenberg implimeted  it in 1952 when he proposed  a very simple model for high energy 
nucleon-nucleon scattering \cite{Heisenberg1952}. He wrote down a massive DBI action to describe 
an interacting ``pion field" that mediates the interaction between nucleons of the following form,
\be
S=\int d^{3+1}x\,l^{-4}\left[1-\sqrt{1+l^4[(\d_\mu\phi)^2+m^2\phi^2]}\right].\label{Heisaction}
\ee

Remarkably, Heisenberg's model reproduces the saturation of the 
the Froissart bound \cite{Froissart:1961ux,Lukaszuk:1967zz},
\be
\sigma_{\rm tot}(s)\leq C\ln^2\frac{s}{s_0}\;,\;\;\; C\leq \frac{\pi}{m_1^2}\;,
\ee
even though the model was proposed not just before Froissart, but even before QCD. In the model, Heisenberg first dimensionally 
reduced the Lagrangian to 1+1 dimensions (time and direction of propagation of the pion field ) and then 
discovered  a shock wave solution of the equation of motion of this model defined perturbatively. 
An analysis of the Heisenberg model, uniqueness properties,
and its generalizations was done in \cite{Nastase:2015ixa}.

Recently there has been a very important development in deciphering the space of field theories in 1+1 dimensions 
in the form of what is referred to as the $T\bar T$ deformations\cite{Zamolodchikov:2004ce,Smirnov:2016lqw}. 
In particular the Lagrangian density for the deformations of canonical scalar action with a potential $V$,  was determined in 
\cite{Cavaglia:2016oda,Bonelli:2018kik,Rosenhaus:2019utc}.
The interest in these deformations comes from the fact that they 
are among very few exactly known quantum deformations. 
While new deformations of non-integrable models were also considered,
the  original motivation to study  these deformations was to explore the subspace of field theories in two 
dimensions which are  integrable. $T\bar T$ deformations have been explored in relation to holography, 
gravity and string theories. In addition  they have served as laboratories to investigate  various aspects 
of field theory, for instance in \cite{Guica:2017lia,Aharony:2018bad,Cardy:2018sdv,Datta:2018thy,Taylor:2018xcy,Brennan:2019azg}.
  
Key  players of integrable field theories are solitons, anti-solitons and their breather bound states. Thus, 
it is important to study the properties of the latter in the $T\bar T$ deformed theories.  This is the task taken in this paper. 
We determine explicit soliton solutions of the $T\bar T$  deformed theories as well as solitons of the Heisenberg model. 
In particular we prove that the classical mass of the soliton of the deformed theory is the same as the mass 
of the soliton  of the undeformed  theory. 
In route to these results we also write down the soliton solution of the Heisenberg model. 
We further write down explicit shockwave solutions of both theories. 
 
The application of these solutions to the Heisenberg model of nucleon nucleon scattering 
in 3+1 dimensions  will be done in a separate publication\cite{Nastase:2020}.

The paper is organized as follows. In section 2 we define the Lagrangians, (generalized) Heisenberg and $T\bar T $ deformations.
In section 3, considering the Heisenberg and
$T\bar T$ deformed scalar actions, we find soliton and single shockwave solutions for them. In section 4 we consider 
perturbative shockwave solutions for them, and in section 5 we conclude. 
%In Appendix A we present an alternative, perturbative, 
%way to constructing soliton solutions and in
 In appendix A we consider extensions of the pure DBI action and Heisenberg actions 
to some cases with arbitrary powers.

\section{$T\bar T$ deformed scalar fields and  Heisenberg model: Actions and equations of motion}

The $T\bar T$ deformation of a Lagrangian, proposed by Zamolodchikov \cite{Zamolodchikov:2004ce,Smirnov:2016lqw}
is that, for a 1+1 dimensional theory Wick rotated to Euclidean space, the deformation with parameter $\lambda$ is
\be
\frac{\d {\cal L}}{\d \lambda}=-4[T^\lambda_{zz}T^\lambda_{\bar z\bar z}-(T^\lambda_{z\bar z})^2].\label{equ}
\ee
where  ${\cal L}(\lambda=0)$  is the original undeformed  Lagrangian 
density, $T_{\mu\nu}^\lambda$ 
  are the components of the energy-momentum
tensor   of the theory deformed by $\lambda$ and 
the composite operator on the right-hand side is  defined  via point
splitting.

The interesting property of  this deformation is that it is of a full quantum theory, and 
 all objects are renormalized and UV finite.  For our purposes we will treat this as simply a certain (effective) classical 
 Lagrangian. 

One can solve the above equation by expanding  in a series in $\lambda$, if we give a 
starting point (an unperturbed Lagrangian
${\cal L}_0$) and a perturbation parameter $\lambda$. It follows from   (\ref{equ}),  that the first order Lagrangian density, 
${\cal L}_1$ is given by the unperturbed determinant of the energy-momentum tensor, $\det T_{\mu\nu}$,
above, etc. 

Consider as a starting point a canonical real scalar field  with a potential $V$ (in Euclidean space), 
\be
{\cal L}_0=\frac{1}{2}(\d_\mu\phi)^2+V\;,
\ee
and a perturbation parameter $\lambda$.

Then, at first one obtained a complicated expression, 
with an infinite series of complicated hypergeometric functions (see eq. 6.34 in \cite{Cavaglia:2016oda}), but the series can be summed, 
to obtain a simple 
expression\cite{Bonelli:2018kik},  giving
\bea
{\cal L}(\lambda,X)&=&\frac{-(1-2\lambda V)+\sqrt{1+2\bar\lambda (\d_\mu \phi)^2}}{2\bar\lambda}\cr
&\equiv& \tilde V(\phi)+\frac{1}{2\bar\lambda}\sqrt{1+2\bar\lambda(\d_\mu\phi)^2}\;,\label{Lagra}
\eea
where we can denote $X=(\d_\mu\phi)^2$, and we have defined
\be
\bar \lambda\equiv \lambda(1-\lambda V).
\ee

Note that when $\lambda V\rightarrow 0$, $\tilde V\rightarrow V-1/(2\lambda)$, which cancels the a 
constant coming from the square root, and gives the 
unperturbed potential. 
This can then also be written as in \cite{Rosenhaus:2019utc} (since, in Euclidean 
complex coordinates, $4\d \phi \bar \d \phi=(\d_\mu \phi)^2$),
\be
{\cal L}_E= \frac{V}{1-\lambda V}+\frac{-1+\sqrt{1+8\bar \lambda \d \phi \bar \d \phi}}{2\bar \lambda}.\label{LagrangE}
\ee

This is the Lagrangian in Euclidean space. Going back to the Minkowski signature, we obtain 
\be
{\cal L}_M=-\frac{V}{1-\lambda V}+\frac{1-\sqrt{1+2\bar \lambda \d_\mu \phi \d^\mu \phi}}{2\bar \lambda}
= - \tilde V -\frac{\sqrt{1+2\bar \lambda \d_\mu \phi \d^\mu \phi}}{2\bar \lambda}\;,\label{Lagrang}
\ee
with $\d_\mu \phi \d^\mu \phi =-(\dot\phi)^2+(\phi')^2\equiv X$ and 
\be
\tilde V\equiv -\frac{1-2\lambda V}{2\lambda(1-\lambda V)}.
\ee 

It is thus obvious that,  depending on the potential and $\lambda$, 
$\tilde V$ can be singular  if there is a value of $\phi$ for which  $V(\phi)=\frac{1}{\lambda}$.
Notice that when we expand $\tilde V$ to leading  order in $\lambda$, namely, 
$\lambda X<<1$ and $\lambda V<<1$, we get $\tilde V\sim  V-\frac{1}{2\lambda}$. But   
when we expand the full  Lagrangian density,    we indeed get $V$ and the term of order $\lambda^{-1}$ cancels out as can be seen from  
\be
{\cal L}_M=\frac12[(\dot\phi)^2-(\phi')^2] - V +\lambda\left [\frac14[(\dot\phi)^2-(\phi')^2]^2 - V^2\right ] + {\cal O}(\lambda^2).
\ee

\begin{figure}[t!] \centering 
\includegraphics[width=0.60\textwidth]{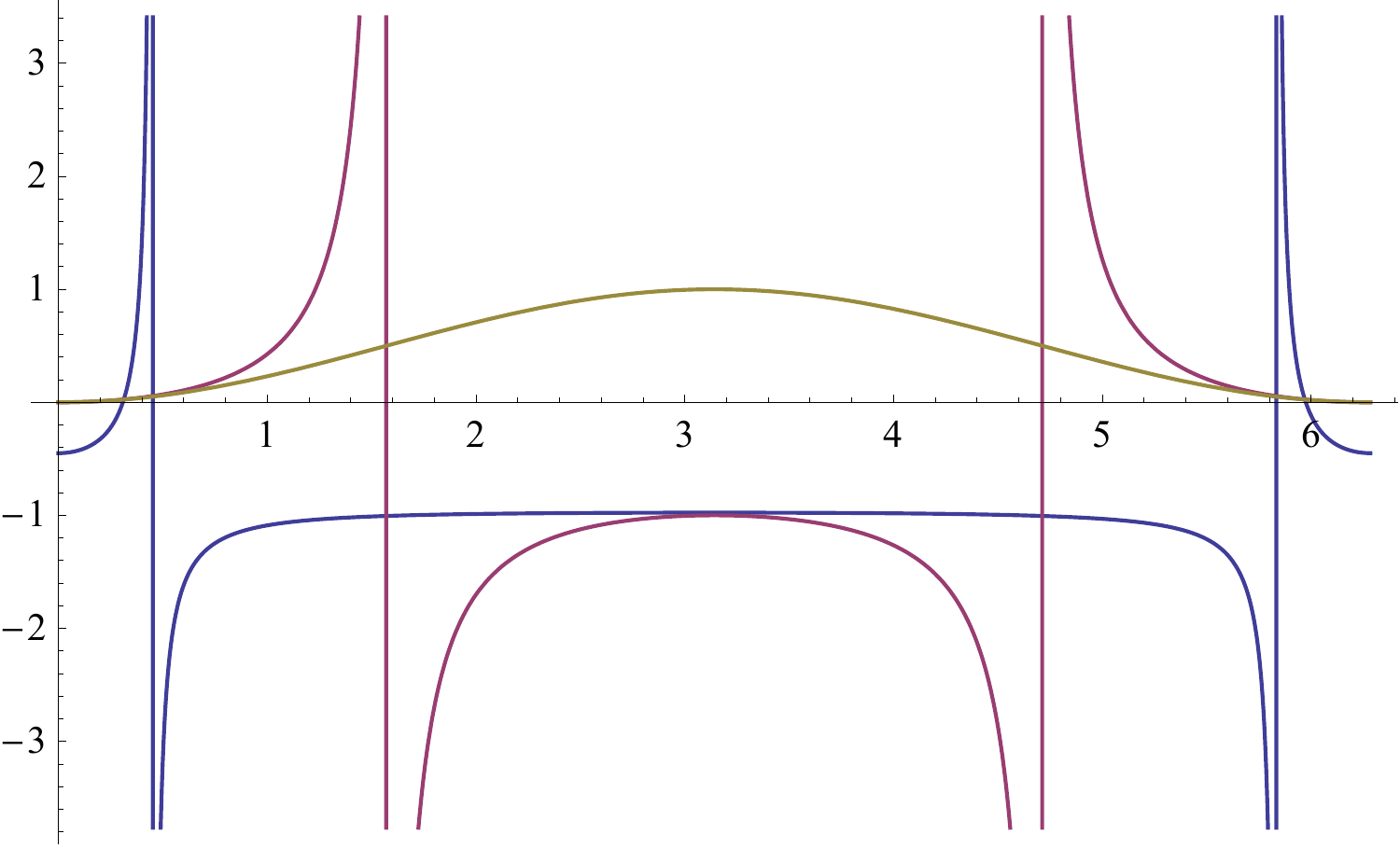}
\caption{\label{fig:tildeV} The  potential of the deformed sine-Gordon model with $\mu=\beta=1$  
for $\lambda=0$. The undeformed case is in brown, $\lambda=1$ is in blue and  $\lambda=10$ is in red.    }
\end{figure}

In Fig.\ref{fig:tildeV} we draw the deformed potential $\tilde V-\frac{1}{2\lambda}$ for the sine-Gordon 
model, for which the undeformed potential is given in (\ref{sG}). %(\ref{potentialsG}).

%%%%%%%%%%%%%%%%%%%

The other Lagrangian we will be interested in is the Heisenberg Lagrangian, with a generic potential 
$V(\phi)$ inside the square root, and dimensionally reduced to 1+1 dimensions, so 
\be
{\cal L}=l^{-4}\left[1-\sqrt{1+l^4[(\d_\mu\phi)^2+2V(\phi)]}\right].\label{Heisenberg}
\ee
When expanding in small $l^4[(\d_\mu\phi)^2+2V(\phi)$ the lagrangian density takes the form
\be
-\frac12 (\d_\mu\phi)^2- V(\phi)  +\frac{l^4}{8}((\d_\mu\phi)^2 + 2V)^2  + {\cal O}(l^8)
\ee

%We now examine whether these Lagrangians (for various possible potentials $V(\phi)$)  
%%%%%%%%%%%%%%%%%%%%%%%%%%%%%%%%%%%%%%%%%%%%

\section{ Soliton and breather solutions}

We now examine whether these $T\bar T$ and Heisenberg Lagrangians (for various possible potentials $V(\phi)$)
admit soliton, breather and shockwave solutions. We first briefly remind the reader the form of these solutions 
in the ordinary sine-Gordon model\footnote{ The deformed sine-Gordon action has been discussed 
in several publication for instance \cite{Smirnov:2016lqw,Cavaglia:2016oda}.}. 
We then analyze the Heisenberg model with a generic potential 
and then the $T\bar T$ deformed action. 

\subsection{The sine-Gordon model} 
 
We start with a review of an un-deformed model with solitons in 1+1 dimensions\footnote{For a review of 
these issues see for instance \cite{Frishman:2010zz}.}.
 
The most  well known  1+1 dimensions scalar field theory that admits soliton solutions is the 
sine-Gordon model  which is defined by the  potential 
\be
V_{sG}= -\frac{\mu^2}{\beta}\left [\cos( \beta \phi) -1 \right ].\label{sG}
\ee

The corresponding equation of motion is
\be
\phi^{\prime\prime}-\ddot\phi -\pa_\phi V=0.
\ee

For static solution we multiply the equation of motion with $\phi^\prime$ to find
\be
\phi^\prime \phi^{\prime\prime}- \phi^\prime\pa_\phi V=0 \rightarrow \qquad
\pa_x\left[ \frac12 (\phi^\prime)^2 - V\right]= 0 \rightarrow  \qquad \frac12 (\phi^\prime)^2 - V= C\;,\label{virial}
\ee
where $C$ is a constant. This is the "virial theorem".
For $C=0$ the equation takes the form
\be
x-x_0 =\frac{\sqrt{\beta}}{\sqrt{2}\mu}\int \frac{d\phi}{\sqrt{1-\cos(\beta\phi)}}\;,
\ee
which yields the  soliton solution
\be
\phi_{\rm sol} = \frac{4}{\beta}\arctan\left [  e^{\pm \mu\sqrt{\beta} (x-x_0)}\right ].
\ee

The mass of the soliton, i.e., its energy is
\bea 
E_{sol}&=&\int dx T_{00}= \int dx \left[ \frac12 (\phi^\prime)^2 +V\right]=\int dx  (\phi^\prime)^2
=\int d\phi \phi^\prime\cr
&=& \int d\phi \sqrt{2 V(\phi)} =\frac{8\mu}{\beta^{3/2}}.\label{masssol}
\eea
 
Next we would like to check  whether the equations of motion admit also time dependent 
exact solutions. Another question seemingly unrelated is, what is the interaction between two solitons and 
between a soliton and an anti-soliton, and whether  there are bound states. In fact the two questions are very related. 

\begin{figure}[t!] \centering 
\includegraphics[width=0.60\textwidth]{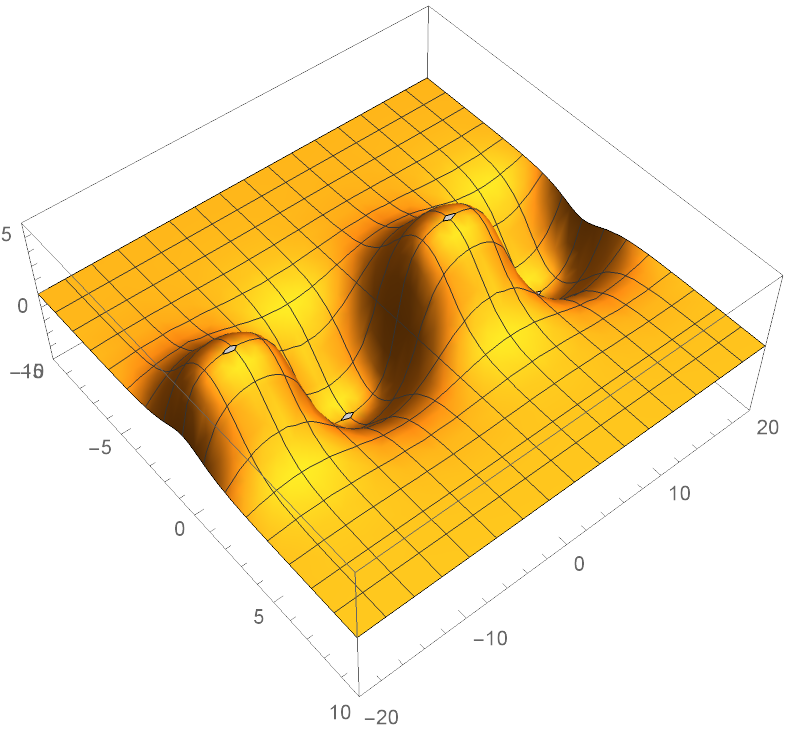}
\caption{\label{figbreather}  The breather solution for $\mu=1,\ \beta=1\tan\psi=2\pi/5$.     }
\end{figure}

It is easy to check that performing the following transformation 
\be
\phi_{\rm sol} = \frac{4}{\beta}\arctan\left [  e^{\pm \mu\sqrt{\beta} (x-x_0)}\right ]\rightarrow \phi_{\rm bre}
=\frac{4}{\beta}\arctan\left [\frac{\eta\sin(wt)}{\cosh(\eta w x)}\right]\;,\label{breather}
\ee
where $\eta =\tan(\psi)= \frac{\sqrt{\mu^2\beta-w^2}}{w}$,
yields a solution of the equation of motion. 
To see that this solution describes a bound state, consider first $w<<\mu\sqrt{\beta}$. When $x\rightarrow -\infty$ we can approximate 
\be\label{breathersol}
 \phi_{\rm bre}\sim\frac{4}{\beta}\arctan\left [e^{ \mu\sqrt{\beta} (x-x_0)  +\log\left(\frac{\mu\sqrt{\beta}}{w}\sin(wt)\right)} \right ]\;,
\ee
which looks like a soliton to the left. Similarly the solution looks like an anti-soliton to the right. 
The breather solition is drawn in Fig.\ref{figbreather}.

Altogether the breather solution describe an oscillating bound state. Indeed we can determine the 
mass of the breather by computing it at $t=0$, for which $\phi^\prime=V=0$, so that 
\be
E_{\rm bre}= \int dx \left [ \frac12 (\dot \phi)^2\right ]= \frac{8(\eta w)^2}{\beta^2}\,\int dx \frac{1}{\cosh^2(\eta w x)}
= 2 M_{\rm sol} \sqrt{ 1- \left ( \frac{w}{\mu\sqrt{\beta}}\right )^2}\;,
\ee
which is obviously smaller than twice the mass of the soliton, namely there is a non-vanishing bin.

\subsection{The Heisenberg model with a generic potential}

Since it is hard to find soliton 
solutions of the $T\bar T$ deformations, we start by first finding solutions of the Heisenberg model modified 
with a generic potential, instead of just a mass term, i.e., the Minkowski space Lagrangian in (\ref{Heisenberg}), namely
\be
{\cal L}_M=1-\sqrt{1+(\d_\mu\phi)^2+2V}\;,
\ee
where we have put $l=1$ for simplicity (to reinstate it, we can just replace $\phi= l^2\tilde \phi$,  $V= l^4\tilde V$ and multiply the
Lagrangian by $l^{-d}$), and 
$(\d_\mu\phi)^2=-\dot\phi^2+\phi'^2$, as before. The Hamiltonian is 
\be
{\cal H}=\dot\phi\frac{\d {\cal L}_M}{\d \dot \phi}-{\cal L}_M= \frac{1+\phi'^2+2V}{\sqrt{1-\dot \phi^2+\phi'^2+2V}}-1.
\ee

For static solutions
\be
{\cal H}= -{\cal L}_M = \sqrt{1 +\phi'^2+2V}-1.
\ee

Consider a static solution, $\dot\phi=0$, so $(\d_\mu \phi)^2=\phi'^2$. Then the equation of motion is 
\be
\frac{d}{dx}\left(\frac{\phi'}{\sqrt{1+\phi'^2+2V}}\right)=\frac{\frac{dV}{d\phi}}{\sqrt{1+\phi'^2+2V}}.
\ee

Multiply it with $\phi'$, so that on the right-hand side we have $(dV/d\phi)(d\phi/dx)=dV/dx$. Then note the identities 
\bea
\frac{d}{dx}\left(\frac{\phi'^2/2}{\sqrt{1+\phi'^2+2V}}\right)&=&\phi'\frac{d}{dx}\left(\frac{\phi'}{\sqrt{1+\phi'^2+2V}}\right)-\frac{\phi'^2}{2}
\frac{d}{dx}\left(\frac{1}{\sqrt{1+\phi'^2+2V}}\right)\cr
\frac{d}{dx}\left(\frac{V}{\sqrt{1+\phi'^2+2V}}\right)&=&\frac{dV/dx}{\sqrt{1+\phi'^2+2V}}+V\frac{d}{dx}\frac{1}{\sqrt{1+\phi'^2+2V}}.
\eea

Subtract them and use the equation of motion above, to obtain 
\bea
\frac{d}{dx}\left(\frac{V-\phi'^2/2}{\sqrt{1+\phi'^2+2V}}\right)&=&(2V+\phi'^2)\frac{d}{dx}\left(\frac{1}{2\sqrt{1+\phi'^2+2V}}\right)\cr
&=&\frac{d}{dx}\left(\frac{V+\phi'^2/2}{\sqrt{1+\phi'^2+2V}}\right)-\frac{d}{dx}\sqrt{1+\phi'^2+2V}\;,
\eea
which finally implies the conservation equation 
\be
\frac{d}{dx}\left(-\frac{1+2V}{\sqrt{1+2V+\phi'^2}}\right)=0\;,\label{conseq}
\ee
with general solution 
\be
\frac{1+2V}{\sqrt{1+2V+\phi'^2}}=C\;,
\ee
for an arbitrary $C$. Solving it for $\phi'^2$, we get
\be
\phi'^2=(1+2V)\left(\frac{1+2V}{C^2}-1\right).
\ee

Then the general solution is written implicitly as 
\be
x-x_0=\int_{\phi(x_0)}^{\phi(x)} \frac{d\phi}{\sqrt{(1+2V)\left[\frac{1+2V}{C^2}-1\right]}}.
\ee

For comparison, consider first the canonical Lagrangian in 1+1 dimensions, ${\cal L}=-\frac{1}{2}(\d_\mu\phi)^2-V$. 
The virial theorem (\ref{virial}) is the conservation equation $\frac{d}{dx}(\phi'^2/2-V)=0$, in the case of {\em finite energy} soliton solutions, 
for which the integration constant $\tilde C$ vanishes, so the solution is 
\be
x-x_0=\int_{\phi(x_0)}^{\phi(x)}\frac{d\phi}{\sqrt{2V(\phi)}}.
\ee

We see that our case is obtained for $C=1$ and in the $l\rightarrow 0$ limit. Keeping $l$ finite (=1 in our convention), but $C=1$, we 
obtain
\be\label{HsGs}
x-x_0=\int_{\phi(x_0)}^{\phi(x)}\frac{d\phi}{\sqrt{2V(1+2V)}}\;,
\ee
corresponding to the equality (deformed virial theorem)
\be
\phi'^2=2V(1+2V).
\ee

Using this deformed virial theorem we can now compute the mass of the soliton
\bea\label{MsHsG}
M_{sol}&=& \int dx {\cal H}= \int dx [\sqrt{1 +\phi'^2+2V}-1] =\int dx 2 V\CR
&=& \int d\phi \frac{2V}{\phi'}= \int d\phi \frac{2V}{\sqrt{2V(1+2V)}}=\int d\phi \frac{\sqrt{2V}}{\sqrt{ (1+2V)}}.\CR
\eea

\begin{figure}[t!] \centering 
\includegraphics[width=0.60\textwidth]{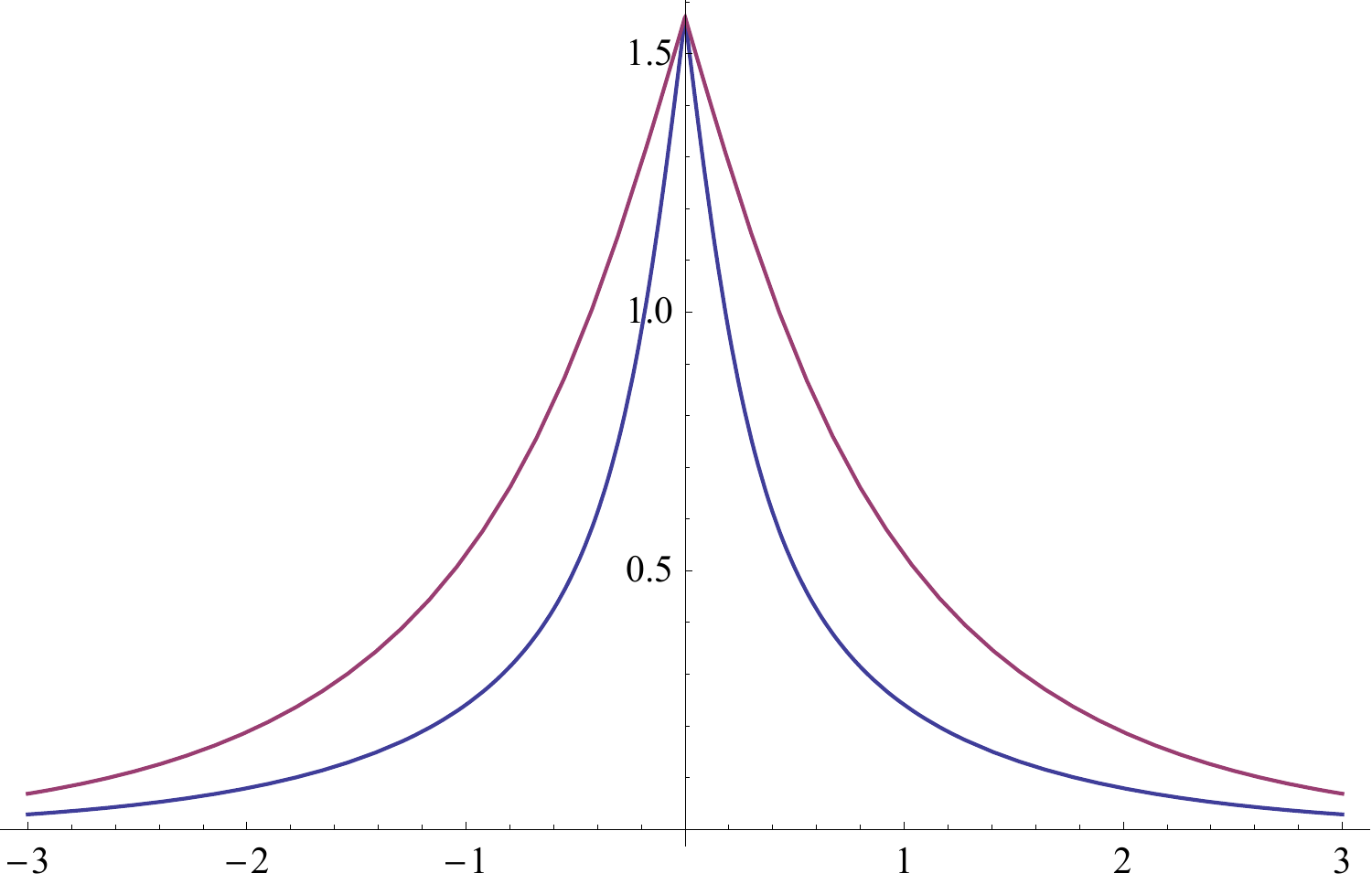}
\caption{\label{fig:Heisenbergsoliton} The soliton (\ref{try}) of the Heisenberg model with a sine-Gordon potential. The colours correspond to: 
A=0.1 for blue and A=10 for purple.  $\frac{\b\phi}{2}$ on the vertical varies from 0 at infinity 
to $\pi/2$ at the origin, corresponding to $x=x_0$.  }
\end{figure}

\begin{figure}[t!] \centering 
\includegraphics[width=0.60\textwidth]{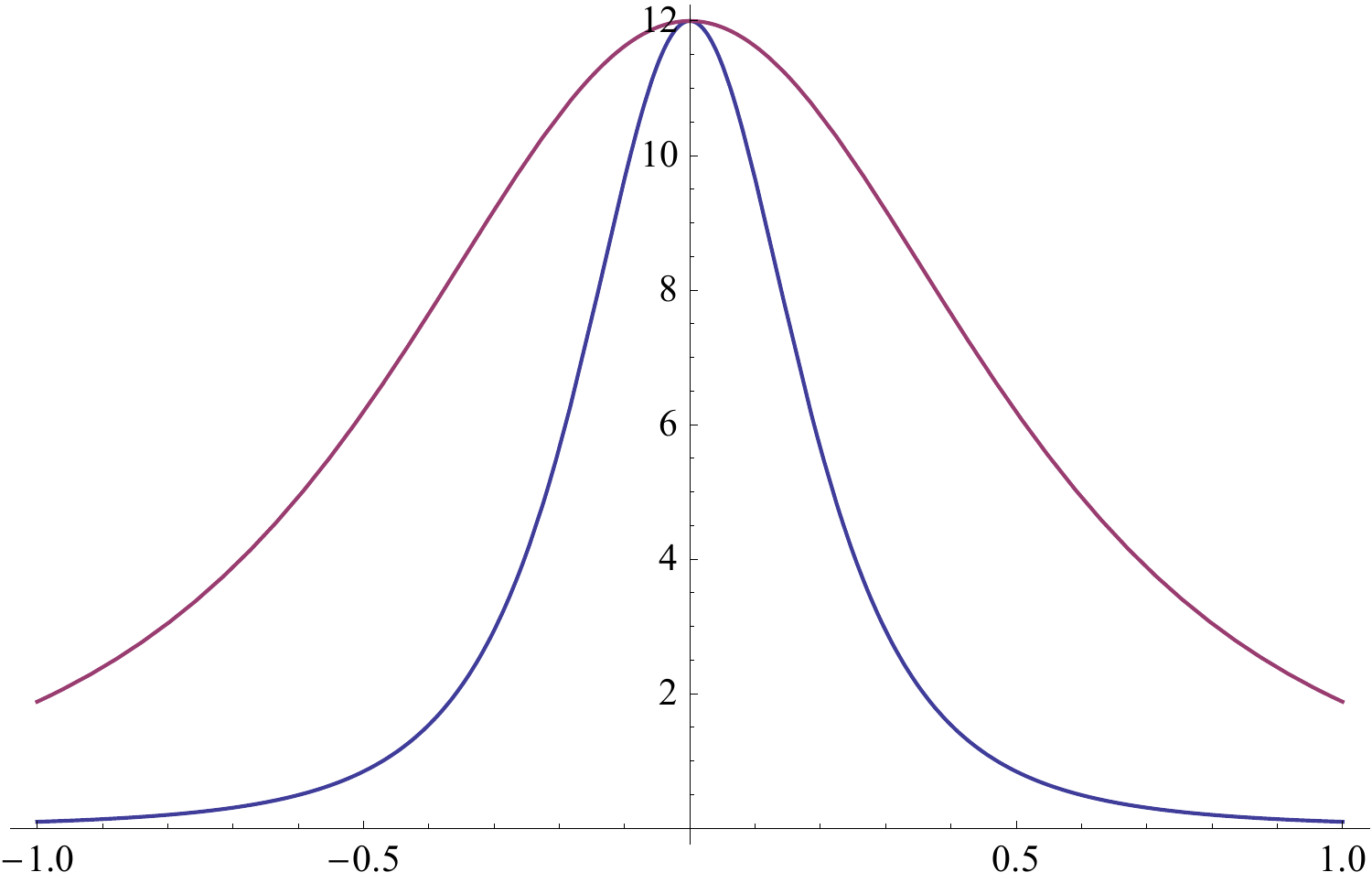}
\caption{\label{fig:HED} $(\phi')^2$ for (\ref{try}) for the Heisenberg model with a sine-Gordon potential. The colors correspond to:
A=0.1 for blue and A=5 for purple.   }
\end{figure}

We can check now what is the soliton solution and what is its corresponding mass for the case of the sine-Gordon potential (\ref{sG}).
Substituting it in (\ref{HsGs}) we get that the solution for $\phi$ reads
\be
x-x_0=\frac{1}{\mu\sqrt{\b}}\tanh^{-1}\left[\frac{\cos\frac{\b\phi}{2}}{\sqrt{1+A\sin^2\frac{\b\phi}{2}}}\right]_\phi^{\frac{\pi}{\b}}\;,\label{tryx}
\ee
or, inverting the formula,
\bea
\phi(x) &=&\frac{2}{\b} \arcsin\left( \sqrt{\frac{1- \tanh^2[-\mu\sqrt{\b}(x- x_0)]}{1+A\tanh^2[-\mu\sqrt{\b}(x- x_0)]} }\right )\cr
&=&\frac{2}{\b} \arctan\left( \sqrt{\frac{1- \tanh^2[-\mu\sqrt{\b}(x- x_0)]}{(1+A)\tanh^2[-\mu\sqrt{\b}(x- x_0)]} }\right)\;,\label{try}
\eea
where  $A =\frac{4 l^2 \mu^2}{\beta}$. However, from (\ref{tryx})
note that $\tanh[-\mu\sqrt{\b}(x-x_0)]$ takes values between $+1$ (for $x=-\infty$)
and $-1$ (for $x=+\infty$), corresponding to $\frac{\b\phi}{2}$ taking values between 0 (for $x=-\infty$) and $\pi$ (for $x=+\infty$), 
while $x=x_0$ is mapped to $\frac{\b\phi}{2}=\frac{\pi}{2}$, just 
like the topological sine-Gordon soliton, whereas $\arcsin$ and $\arctan$ have range $[-\pi/2,\pi/2]$, so (\ref{try}) produces
a result in $[0,\pi/2]$. Then, the topological soliton that is a deformation of the sine-Gordon soliton 
(and so must go between $0$ at $-\infty$ and $\frac{2}{\b}\pi$ at $+\infty$) is defined as (\ref{try})
for $x\leq x_0$, and as $\pi-$(\ref{try}) for $x\geq x_0$, which ensures also that $\phi'$ is continuous at $x_0$. 
On the other hand, (\ref{try}) can be taken to be a solution for all $x\in \mathbb{R}$.
%and we have shifted $x_0$ to $\tilde x_0$ by the value of the integral at $\phi=0$.
%However, since that value is actually infinite, if we integrate to $\phi=0$, the solution must be understood to be valid only on the half
%line, from $x=0$ to $x=\infty$ only, and for the solution from $-\infty$ to 0, we should cut and glue another solution at $\phi=0$. 
%This solution (\ref{try}),  defined on the full line, in the $Arctan$ form  is drawn in Fig.\ref{fig:solsG} and the corresponding $(\phi')^2$ in Fig.\ref{fig:dphisquare}.
The solution defined in (\ref{try}) is drawn in Fig.\ref{fig:Heisenbergsoliton} and the corresponding $(\phi')^2$ in 
Fig.\ref{fig:HED}.

Note also that the limit $l\rightarrow 0$ of the deformed topological soliton, 
which should give back the undeformed soliton, means $A\rightarrow 0$, 
in which case we can simplify (\ref{try}) using the fact that $\tanh^{-1}(\cos 2x)=y$ implies $x=\arctan  e^{\pm y}$. Then, indeed, the 
implicit solution on the first line of (\ref{try}) becomes $\phi=\frac{4}{\b}\arctan[e^{\pm \mu \sqrt{\b}(x-x_0)}]$, which is just the 
undeformed sine-Gordon soliton.

We can also determine the mass associated with this solution and check in particular if it is finite, and hence indeed a soliton.
We find
\be
M_{solsG}= \int d\phi \frac{\sqrt{2V}}{\sqrt{ (1+2V)}}=\frac{4}{l\beta}\arctan\left(\frac{2l\mu}{\sqrt{\beta}}  \right)\;,
\ee
where we used that $\frac{\beta\phi}{2}$ changes in the interval $[0,\pi]$ 
(note that the variation of $\frac{\b\phi}{2}$ in Fig.\ref{fig:Heisenbergsoliton} is from 0 to $\pi/2$ to 0, but as we said,
the topological soliton corresponds to $\pi-$(\ref{try}) on the positive real axis).
%which is one quarter of this, so we would need to glue several solutions to fill the interval).
This is indeed a finite result, and in the limit $l\mu\rightarrow 0$, it goes back to the mass of the ordinary sine-Gordon soliton.

\subsubsection{Pure DBI action}

As another comparison, for the pure DBI action, at $V(\phi)=0$, we obtain
\be
x-x_0=\int_{\phi(x_0)}^{\phi(x)}\frac{2d \phi}{\sqrt{\frac{1}{C^2}-1}}=K(\phi(x)-\phi(x_0))\;,
\ee
where $K$ is another constant, related to $C$. Note that we can glue together two such solutions to obtain the solution to the 
Poisson equation in one dimension, 
\be
\phi(x)=\phi(x_0)+K^{-1}|x-x_0|.
\ee

Note that if we boost this solution, we obtain 
\be
\phi(x)=\phi(x'_0)+K^{-1}\gamma|x-x'_0-vt|\;,
\ee
or, in the ultrarelativistic limit $v\rightarrow 1$, with $\phi(x'_0)=0$ (since it becomes anyway negligible with respect to the second term 
in the $\gamma\rightarrow\infty$ limit) and $K^{-1}\gamma\equiv \tilde K$, 
\be
\phi(x)=\tilde K|x^-|.
\ee
This is clearly not an soliton since it does not have finite mass.

\subsection{Soliton solutions of  the $T\bar T$ deformed scalar field}

We are finally in a position to write the same for the $T\bar T$ deformed Lagrangian. 

For a static solution, the equation of motion of the Lagrangian (\ref{Lagrang}) is 
\bea
&&\frac{d}{dx}\left(\frac{\phi'}{\sqrt{1+2\lambda(1-\lambda V)\phi'^2}}\right)\cr
&=&
-\frac{\lambda}{2(1-\lambda V)}\frac{\frac{dV}{d\phi}\phi'^2}{\sqrt{1+2\lambda(1-\lambda V)\phi'^2}}+\frac{\lambda\frac{dV}{d\phi}}{2
(1-\lambda V)^2}\sqrt{1+2\lambda (1-\lambda V)\phi'^2}+\frac{d\tilde V}{d\phi}.
\eea

As before, we multiply the equation by $\phi'$, and find on the right-hand side $(dV/d\phi)(d\phi/dx)=dV/dx$ and $(d\tilde V/d\phi)
(d\phi/dx)=d\tilde V/dx$. We then note the identities 
\bea
&&\frac{d}{dx}\left(\frac{\phi'^2/2}{\sqrt{1+2\lambda (1-\lambda V)\phi'^2}}\right)= \phi' \frac{d}{dx}\left(\frac{\phi'}
{\sqrt{1+2\lambda (1-\lambda V)\phi'^2}}\right)-\frac{\phi'^2}{2}\frac{d}{dx}\left(\frac{1}{\sqrt{1+2\lambda (1-\lambda V)\phi'^2}}\right)\cr
&&\frac{d}{dx}\left(-\frac{\lambda V}{2(1-\lambda V)}\frac{\phi'^2}{\sqrt{1+2\lambda (1-\lambda V)\phi'^2}}+\frac{\lambda V}{2(1-\lambda V)^2}
\sqrt{1+2\lambda (1-\lambda V)\phi'^2}+\tilde V\right)\cr
&=& -\frac{\lambda}{2(1-\lambda V)}\frac{\frac{dV}{dx}\phi'^2}{\sqrt{1+2\lambda(1-\lambda V)\phi'^2}}+\frac{\lambda\frac{dV}{dx}}{2
(1-\lambda V)^2}\sqrt{1+2\lambda (1-\lambda V)\phi'^2}+\frac{d\tilde V}{dx}\cr
&&+V\frac{d}{dx}\left[-\frac{\lambda}{2(1-\lambda V)}\frac{\phi'^2}{\sqrt{1+2\lambda (1-\lambda V)\phi'^2}}+\frac{\lambda}{2(1-\lambda V)^2}
\sqrt{1+2\lambda (1-\lambda V)\phi'^2}\right]\;,
\eea
and by subtracting them and using the equations of motion above, we get 
\bea
&&\frac{d}{dx}\left(-\frac{\lambda V}{2(1-\lambda V)}\frac{\phi'^2}{\sqrt{1+2\lambda (1-\lambda V)\phi'^2}}
+\frac{\lambda V \sqrt{1+2\lambda (1-\lambda V)\phi'^2}}{2(1-\lambda V)^2}+\tilde V -\frac{\phi'^2/2}
{\sqrt{1+2\lambda (1-\lambda V)\phi'^2}}\right)\cr
&=&\frac{\phi'^2}{2}\frac{d}{dx}\frac{1}{\sqrt{1+2\lambda (1-\lambda V)\phi'^2}}+V\frac{d}{dx}\left[-\frac{\lambda \phi'^2}
{2(1-\lambda V)\sqrt{1+2\lambda (1-\lambda V)\phi'^2}}+\frac{\lambda \sqrt{1+2\lambda (1-\lambda V)\phi'^2}}{2(1-\lambda V)^2}\right]\cr
&=&\frac{d}{dx}\left(\frac{\phi'^2}{2}\frac{1}{\sqrt{1+2\lambda (1-\lambda V)\phi'^2}}
+V\left[-\frac{\lambda \phi'^2}
{2(1-\lambda V)\sqrt{1+2\lambda (1-\lambda V)\phi'^2}}
+\frac{\lambda \sqrt{1+2\lambda (1-\lambda V)\phi'^2}}{2(1-\lambda V)^2}\right]\right)\cr
&&-\left(\frac{d\phi'^2}{dx}\frac{\d}{\d \phi'^2}+\frac{d V}{dx}\frac{\d }{\d V}\right)\left(
\frac{\sqrt{1+2\lambda (1-\lambda V)\phi'^2}}{2\lambda (1-\lambda V)}\right)\cr
&=&\frac{d}{dx}\left(\frac{\phi'^2}{2}\frac{1}{\sqrt{1+2\lambda (1-\lambda V)\phi'^2}}
+V\left[-\frac{\lambda \phi'^2}
{2(1-\lambda V)\sqrt{1+2\lambda (1-\lambda V)\phi'^2}}+\frac{\lambda
 \sqrt{1+2\lambda (1-\lambda V)\phi'^2}}{2(1-\lambda V)^2}\right]\right.\cr
&&\left.-\frac{\sqrt{1+2\lambda (1-\lambda V)\phi'^2}}{2\lambda (1-\lambda V)}\right)\;,
\eea
so that finally we have the conservation equation
\be
\frac{d}{dx}\left[\frac{1}{2\lambda(1-\lambda V)\sqrt{1+2\lambda (1-\lambda V)\phi'^2}}+\tilde V\right]=0.
\ee

It is solved by 
\be\label{solTbT}
\frac{1}{2\lambda(1-\lambda V)\sqrt{1+2\lambda (1-\lambda V)\phi'^2}}+\tilde V=C\;,
\ee
for a general $C$. 

The implicit general solution for $\phi(x)$ is then 
\be
x-x_0=\int_{\phi(x_0)}^{\phi(x)}\frac{d\phi \sqrt{2\lambda (1-\lambda V)}}{\sqrt{\frac{1}{[2\lambda (1-\lambda V)(C-\tilde V)]^2}-1}}.
\ee

Consider the case $C=0$, like in the case of the canonical scalar with potential. Then then solution becomes much simpler:
\be
x-x_0=\int_{\phi(x_0)}^{\phi(x)} \frac{d\phi|1-2\lambda V|}{\sqrt{2V}}.\label{defsol}
\ee

%An  alternative method of deriving the soliton solution based on perturbative expansion in $\bar\lambda X<<1$ is described in appendix A.
%We would like to check for what potential does the $T\bar T$ deformed admit a soliton solution.
%%%%%%%%%%%%%

We can now determine the mass of the soliton of the deformed $T\bar T$ action.
%Recall that for the undeformed  soliton the mass is given by (\ref{masssol}) and is $ =\frac{8\mu}{\beta^{3/2}} $.
We first calculate the Hamiltonian density on the static solution (soliton),
\be
{\cal H}=T_{00}=\frac{\pa {\cal L}}{\pa \dot \phi} \dot \phi - {\cal L}=-{\cal L}.
\ee

Using the condition for a soliton solution of the equation of motion (\ref{defsol}) inside 
${\cal H}=-{\cal L}(\dot \phi=0)$,  we find that the {\em on-shell} Hamiltonian is 
%( for details see Appendix C)
\be
\phi'=\frac{\sqrt{2V}}{|1-2\lambda V|}\Rightarrow 
{\cal H}=\frac{2V}{1-2\lambda V}\;,\label{Hlambda}
\ee
where we have substituted the expression for $\phi$ {\em on the solution} in the Lagrangian.

Recall now the derivation of the soliton mass of the undeformed theory 
\bea 
M_{s}&=& \int_{-\infty}^{+\infty} dx\, T_{00}= \int_{-\infty}^{+\infty} dx\left[ \frac{1}{2} (\phi')^2 + V\right]
=\int_{-\infty}^{+\infty} dx \, 2 V\cr
&=&\int_{-\infty}^{+\infty} dx\, \phi' \sqrt{2V}=\int_0^\pi d\phi \sqrt{2V},
\eea

For the deformed solution we use the above on-shell Hamiltonian, and from  (\ref{defsol}) we replace $\sqrt{2V}/(1-2\lambda V)$ 
with $\phi'$,  so the mass is given by 
\be
M_{s,T\bar T}= \int_{-\infty}^{+\infty} dx\,  {\cal H} = \int_{-\infty}^{+\infty} dx\, \frac{2V}{1 -2 \lambda V}=\int_0^\pi d\phi \,
\sqrt{2V}=M_{s,undef.}
 \;,
\ee
so  we find  a general statement that: 

{\bf there is no correction to the soliton mass at all! }

This is, no doubt, due to the special nature of the $T\bar T$ deformation. Note also that 
we have been able to calculate the mass of the deformed soliton, even though we do not have an explicit analytic expression 
for $\phi(x)$, only an implicit one, because we have traded the $x$ integral for the $\phi$ integral, and we know the values 
of $\phi$ at $x=\pm \infty$, where the soliton is unmodified. 

The reason that we know the solution for $\phi$ near $x=\pm \infty$, namely the un-deformed solution, is the following. 
Consider the implicit deformed solution (\ref{defsol}) near $x=\pm \infty$. Assume that the undeformed solution $\phi$ at $x=\pm \infty$ 
is finite, which is true for most cases of solitons. Moreover, since the mass of these solitons must be finite, then $V(\phi)$ 
must be finite (actually, must be going to zero) near $\phi(x=\pm \infty)$. Then the deformation term in (\ref{defsol}), 
$-\lambda \int d\phi \sqrt{2V}$, is also finite, so can be ignored with respect to the first term, which is infinite (because the left-hand 
side of (\ref{defsol}) is infinite). It follows then that the soliton solution near $x=\pm \infty$ is the undeformed soliton one.

%%%%%%%%%%%%%%%%%%%
\subsubsection{Deformation of sine-Gordon} 

For the  $T\bar T$ deformation of the sine-Gordon potential (\ref{sG}), 
%where we identify
%\be
%\frac{\mu^2}{\b}\equiv \frac{1}{2\lambda}\;,
%\ee
we obtain the following  solution 
\bea
\frac{\mu}{\sqrt{\b}}(x-x_0)&=&\int_{\phi(x_0)}^{\phi(x)}d\phi\frac{|1-4\frac{\lambda \mu^2}{\b}\sin^2\b\phi/2|}{2|\sin \b\phi/2|}\cr
&=&\pm \frac{1}{\b}\left.\left[\ln\left(\tan \frac{\b\phi}{4}
\right)+4\frac{\lambda \mu^2}{\b}\cos\frac{\b\phi}{2}\right]\right|_{\phi(x_0)}^{\phi(x)}\;,
\eea
or
\be
\pm \mu\sqrt{\b}(x-x_0)=\left.\left[\ln\left(\tan\frac{\b\phi}{4}\right)+4\frac{\lambda \mu^2}{\b}
\cos\frac{\b\phi}{2}\right]\right|_{\phi(x_0)}^{\phi(x)}.
\ee

We note that near $\phi=0$ or $\phi=+\infty$, corresponding to $\pm x=-\infty$ or $+\infty$, respectively, the solution is the 
sine-Gordon soliton, as expected from the general analysis above. 
It is only near $x=x_0$ that the soliton is changed. So we can think of it as a modification of the 
sine-Gordon soliton. Moreover, the limit $\lambda\rightarrow 0$ exactly gives the sine-Gordon soliton, as it should, since 
in this limit the action goes to the sine-Gordon soliton action as well. 

Following the result in the general case, the mass of the soliton of the deformed sine-Gordon action is unmodified, namely 
\be
M_{s, T\bar T sg}=\frac{8\mu}{\beta^{3/2}}
\ee

%We can now determine the mass of the soliton of the deformed sine-Gordon model.
%Recall that for the undeformed  soliton the mass is given by (\ref{masssol}) and is $ =\frac{8\mu}{\beta^{3/2}} $.
%We first calculate the Hamiltonian density on the static solution (soliton),
%\be
%{\cal H}=T_{00}=\frac{\pa {\cal L}}{\pa \dot \phi} \dot \phi - {\cal L}=-{\cal L}.
%\ee

%Using the condition for a soliton solution of the equation of motion (\ref{defsol}),  we find that the {\em on-shell} Hamiltonian is
%\be
%\phi'=\frac{\sqrt{2V}}{|1-2\lambda V|}\Rightarrow 
%{\cal H}=\frac{2V}{1-2\lambda V}\;,\label{Hlambda}
%\ee
%where we have substituted the expression for $\phi$ {\em on the solution} in the Lagrangian. 
%and hence 
%\be
%T_{00}= \tilde V - \frac{1}{4{\bar\lambda}^2 \tilde V}= \tilde V\left (  1-\frac{1}{4{\bar\lambda}^2}\right ) = 2V
%\ee
%Thus surprisingly the Hamiltonian density of the deformed action has the same form as the undeformed one. 
%However, since the solution of the equation of motion for $\phi$ is deformed we find
%\be

%\ee

\begin{figure}[t!] \centering 
\includegraphics[width=0.60\textwidth]{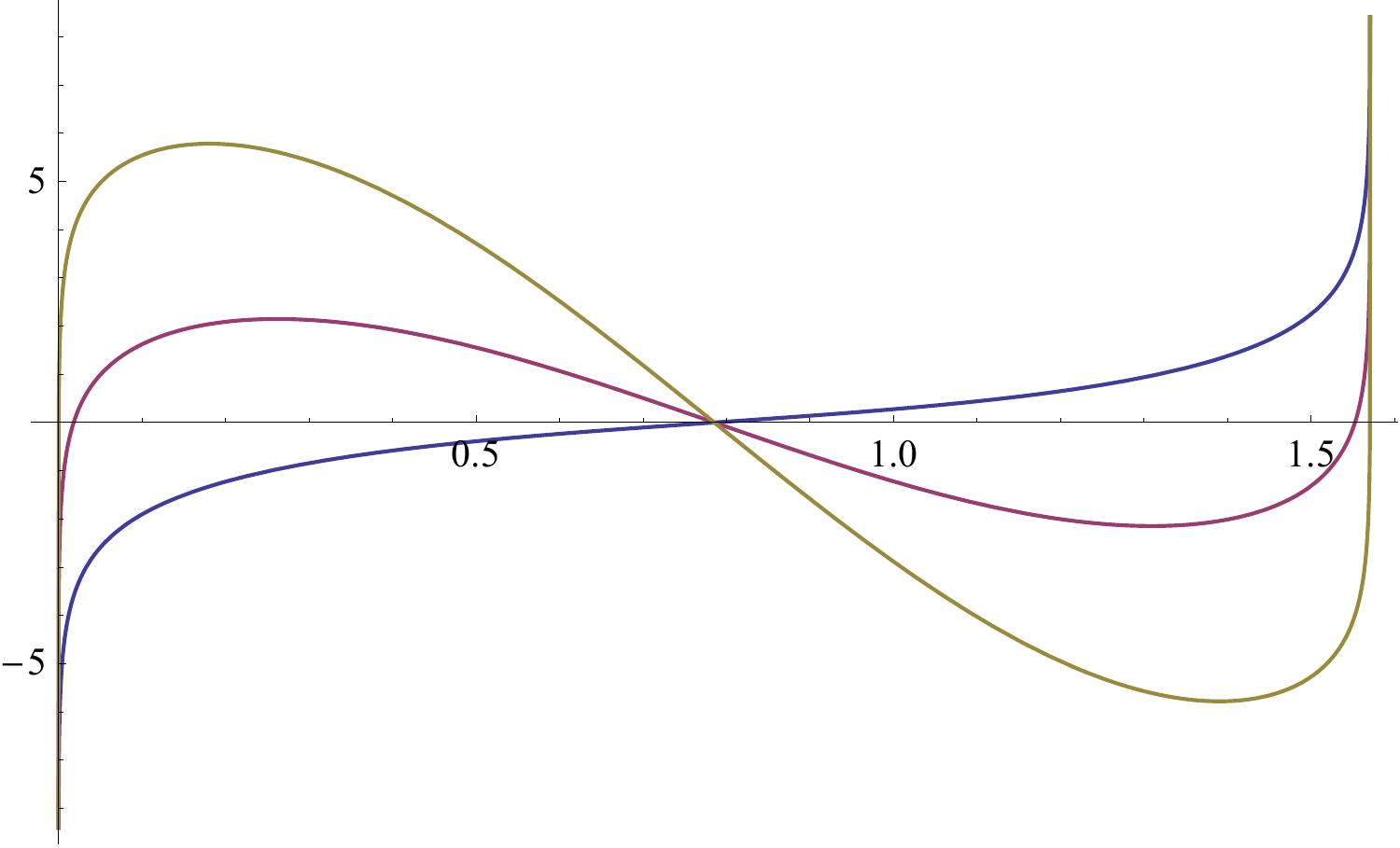}
\caption{\label{fig:tbtsoliton}The  profile of the modified Sine-Gordon 
soliton. $\mu\sqrt{\beta}(x-x_0)$ as a function of $\beta\phi/4 $ for 
$\frac{\lambda \mu^2}{\beta}$=0.1, 1, 2 in blue, purple and brown, respectively.}
\end{figure}

\subsubsection{Higgs-type potential}

We can also consider the Higgs-like potential that gives (in the canonical scalar case) the kink solution, 
\be
V(\phi)=\a(\phi^2-a^2)^2.
\ee

The usual kink solution is 
\be
\phi=a\tanh \left(\pm a\sqrt{2\a}(x-x_0)\right)\;,
\ee
or 
\be
x-x_0=\pm \frac{1}{a\sqrt{2\a}}\tanh^{-1} \frac{\phi}{a}.
\ee

\begin{figure}[t!] \centering 
\includegraphics[width=0.60\textwidth]{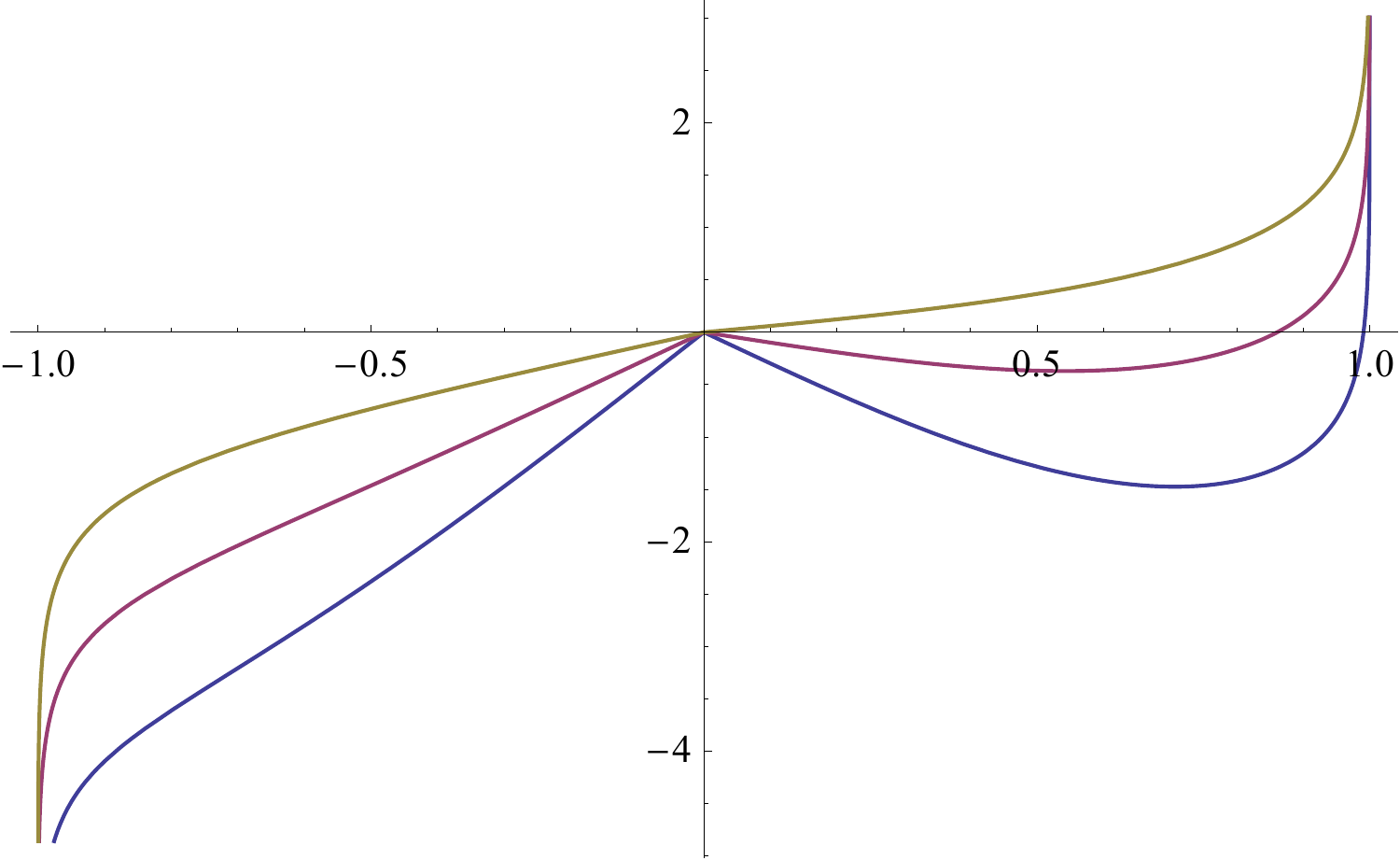}
\caption{\label{fig:Higgsoliton}The  profile of the modified Higgs-like  
soliton. $a\sqrt{2\alpha}(x-x_0)$ as a function of $\phi/a $ for $a=1$ and  
${\lambda \alpha}$=0.1, 0.5, 1 in brown, purple and blue, respectively.}
\end{figure}

In our case, we obtain 
\be
x-x_0=\int \frac{d\phi|1-4\lambda \a(\phi^2-a^2)^2|}{\sqrt{2\a}|a^2-\phi^2|}\;,
\ee
implying
\be
x-x_0=\pm \frac{1}{a\sqrt{2\a}}\left[\tanh^{-1} \frac{\phi}{a}-4\lambda \a a^3 \left|\frac{\phi^3}{3a^3}-\frac{\phi}{a}\right|\right]\;,
\ee
which again can be thought of as a modification of the kink solution, once we realize that we need to restrict to $\phi\in [-a,+a]$, since
$\phi(\pm a)=\pm \infty$. Then near $x=\pm \infty$, the solution is unmodified, as expected from the general analysis, 
while it is modified near $x=x_0$, though this time we 
still have $\phi(x_0)=0$ for the modified kink. The profile of the soliton is shown in Fig.\ref{fig:Higgsoliton}.

If we can neglect the potential $V(\phi)$ altogether, 
for instance if we consider a small $x-x_0$, near a point where $\phi(x_0)$ is also small (so that 
$V(\phi)$ is also small), then we find again (as in the previous subsections) 
a linear solution $x-x_0=K(\phi(x)-\phi(x_0))$, and by gluing two of those, we can again 
find a solution to the Poisson equation in one dimension, the same 
\be
\phi(x)=\phi(x_0)+K^{-1}|x-x_0|
\ee
as before, which after an infinite boost goes over to the same 
\be
\phi(x)=\tilde K |x^-|.
\ee

\subsection{Breathers of the $T\bar T$ system}

The construction of  breather solutions to the sine-Gordon model (\ref{breather}) was based on performing a transformation  of the form 
\be
  \left [ e^{\pm \mu\sqrt{\beta} (x-x_0)} \right ]\rightarrow  \left [\frac{\eta\sin(wt)}{\cosh(\eta w x)}\right].
\ee

This naturally leads us to  conjecture that  the breather solution for the $T\bar T$ deformed sine-Gordon system takes the form
\be
\ln\left [\frac{\eta\sin(wt)}{\cosh(\eta w x)}\right]=\left.\left[\ln\left(\tan\frac{\b\phi}{4}\right)+4\frac{\lambda \mu^2}{\b}
\cos\frac{\b\phi}{2}\right]\right|_{\phi(x_0)}^{\phi(x)}.\label{conjecture}
\ee

For a scalar field that depends on both $x$ and $t$, the equation of motion reads
(remember that $X=(\d_\mu \phi)^2=-\dot \phi^2+\phi'^2$)
\bea
&&-\frac{\lambda}{2(1-\lambda V)}\frac{\frac{dV}{d\phi}X}{\sqrt{1+2\lambda(1-\lambda V)X}}+\frac{\frac{dV}{d\phi}}{2
(1-\lambda V)^2}\sqrt{1+2\lambda (1-\lambda V)X}+\frac{d\tilde V}{d\phi}\cr
&=&\pa_\mu \left(\frac{\pa^\mu\phi}{\sqrt{1+2\lambda(1-\lambda V)X}}\right).
\eea

Normally, we should check our conjectured breather solution (\ref{conjecture}) on the above equation of motion, but that is 
increasingly complicated. Instead, we note that when we differentiate (\ref{conjecture}) with respect to either $t$ or $x$, we obtain 
on the right-hand side
\be
\frac{\b}{2\sin \frac{\b \phi}{2}}(1-2\lambda V(\phi))(\dot \phi\;\;{\rm or}\;\;\phi').
\ee

Then the soliton solution satisfies 
\be
\phi'=\frac{1}{1-2\lambda V(\phi)}\frac{d}{dx}\frac{4}{\b}\arctan \left[e^{\pm \mu\sqrt{\beta} (x-x_0)} \right]\;,%=\pm \mu\sqrt{\b}\;,
\ee
besides satisfying the sine-Gordon equation, $\phi''-\ddot \phi=\mu^2\sin (\b \phi)$. The conjectured breather solution would 
satisfy
\bea
\phi'&=&\frac{1}{1-2\lambda V(\phi)}\frac{d}{dx}\frac{4}{\b}\arctan \left [\frac{\eta\sin(wt)}{\cosh(\eta w x)}\right]\cr
\dot \phi&=&\frac{1}{1-2\lambda V(\phi)}\frac{d}{dt}\frac{4}{\b}\arctan \left [\frac{\eta\sin(wt)}{\cosh(\eta w x)}\right]\;,\label{phiprdot}
\eea
which is true for the usual breather (at $\lambda=0$). Thus in effect we have $\dot \phi\rightarrow \dot \phi/(1-2\lambda V)$
and $\phi'\rightarrow \phi'/(1-2\lambda V)$, resulting in $X=-\dot\phi^2+\phi'^2\rightarrow X/(1-2\lambda V)^2$, {\em both for the soliton 
and for the breather}, which is why our conjectured solution is most likely correct.

\begin{figure}[t!] \centering 
\includegraphics[width=0.60\textwidth]{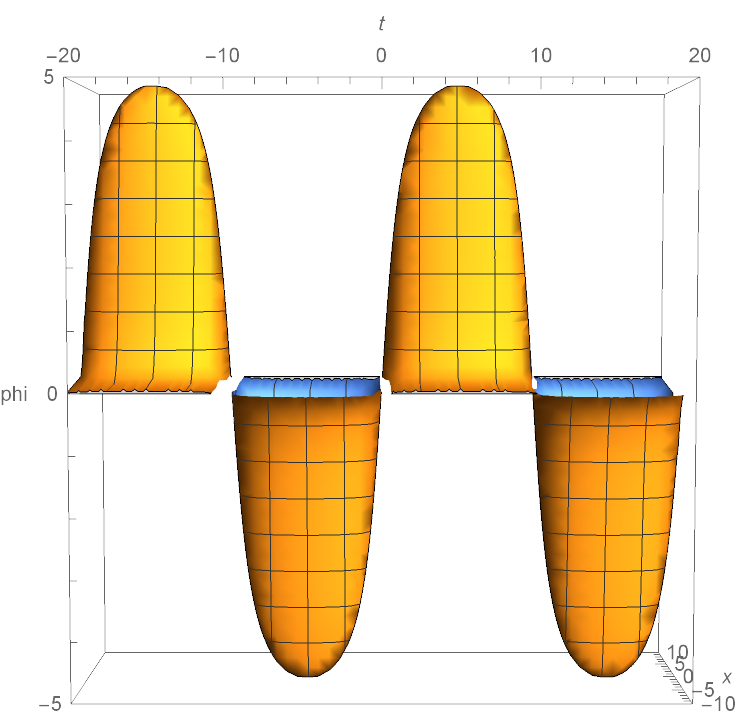}
\caption{\label{figbreatherud}  The undeformed breather solution for $\mu=1,\ \beta=1,\tan\psi=2\pi/5$.}
\end{figure}

\begin{figure}[t!] \centering 
\includegraphics[width=0.60\textwidth]{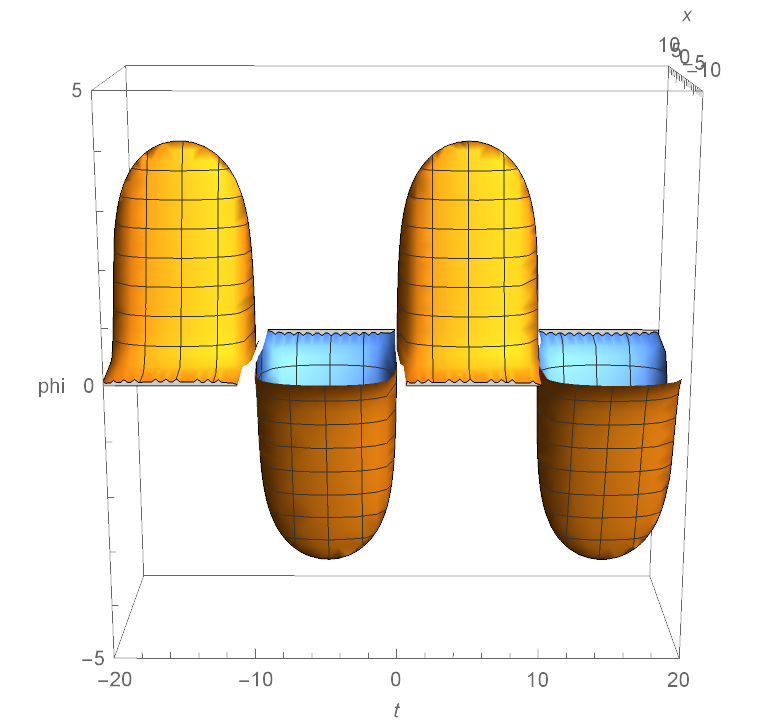}
\caption{\label{figbreatherundefm1b12pi5xlm05}  The deformed breather solution for $\mu=1,\ \beta=1,\tan\psi=2\pi/5,\  \lambda=-1/2 $.    }
\end{figure}

\begin{figure}[t!] \centering 
\includegraphics[width=0.60\textwidth]{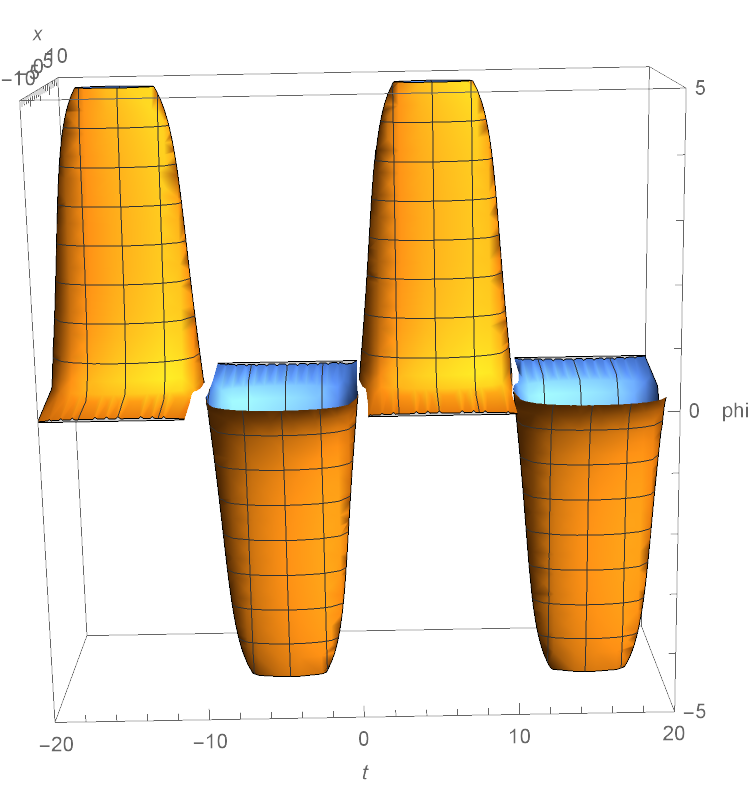}
\caption{\label{figbreatherundefm1b12pi5xl01}  The breather solution for $\mu=1,\ \beta=1,\tan\psi=2\pi/5,\ \lambda=1/10$.     }
\end{figure}

%%%%%%%%%%%%%%%%%%%%%%%%%%%
Similar to the way that we have shown that the breather is indeed a boundstate of a soliton anti-soliton  
(\ref{breathersol}) we can check now the deformed breather for  $w<<\mu\sqrt{\beta}$ and in the limit 
$x\rightarrow -\infty$, which takes the form 
\be
\pm \mu\sqrt{\b}(x-x_0)\log\left(\frac{\mu\sqrt{\beta}}{w}\sin(wt)\right ) =\left.\left[\ln\left(\tan\frac{\b\phi}{4}\right)+4\frac{\lambda \mu^2}{\b}
\cos\frac{\b\phi}{2}\right]\right|_{\phi(x_0)}^{\phi(x)}.
\ee
which indeed looks like the deformed soliton to the left.
In Fig.\ref{figbreatherud} we re-draw   the  breather of the undeformed theory and then 
in Fig.\ref{figbreatherundefm1b12pi5xlm05} and Fig.\ref{figbreatherundefm1b12pi5xl01} the breathers of the same parameters 
with $\lambda=-1/2$ and $\lambda=1/10$, respectively.


%%%%%%%%%%%%%%%%%%%%%%
Furthermore, we can use the same logic as in the undeformed case to calculate the mass of the deformed breather. 
Namely, we first calculate the Hamiltonian, 
\bea
{\cal H}&=& \dot \phi \frac{\d {\cal L}}{\d \dot \phi}-{\cal L}\cr
&=&\frac{(\dot \phi)^2}{\sqrt{1+2\bar \lambda(-\dot\phi^2+\phi'^2)}}+\tilde V +\frac{1+2\bar\lambda(-\dot \phi^2+\phi'^2)}{2\bar\lambda}\;,
\eea
and then we calculate the mass at $t=0$, when we observe that $\phi'(t=0)=0$ and $\phi(t=0)=0$ using (\ref{phiprdot}) and 
(\ref{conjecture}), so also $V(\phi(t=0))=0$, like for the undeformed breather. This implies $\bar \lambda=\lambda$ and 
$\tilde V=-1/(2\lambda)$. Then we find 
\bea
M_{bre,T\bar T}&=&\int_{-\infty}^{+\infty} dx\;{\cal H}(t=0)=\frac{1}{2\lambda}\int_{-\infty}^{+\infty}dx \left(\frac{1}{\sqrt{1-2\lambda \dot \phi^2}}
-1\right)\cr
&=&\int_{-\infty}^{+\infty}dx \sum_{n=0}^\infty\frac{(-1/2)(-1/2-1)...(-1/2-n)}{1\cdot 2\cdot ...\cdot (n+1)}(2\lambda)^n
\dot\phi^{2(n+1)}\cr
&\simeq & \int_{-\infty}^{+\infty}dx \left[\frac{\dot\phi^2}{2}+\frac{3}{8}2\lambda \dot \phi^4+...\right].
\eea

Since, as in the undeformed case, 
\be
\dot \phi^2(t=0)=\left(\frac{4\eta\omega}{\b}\right)^2\frac{1}{\cosh^2(\eta\omega x)}\;,
\ee
doing the integrals we find
\be
M_{bre,T\bar T}=M_{bre}\left[1+\sum_{n=1}^\infty\frac{(-1/2)(-1/2-1)...(-1/2-n)}{1\cdot 2\cdot ...\cdot (n+1)}B^n \sum_{k=0}^n \frac{n!}{
(2k+1)k!(n-k)!}\right]\;,
\ee
where 
\be
B\equiv 2\lambda \left(\frac{4\eta \omega}{\b}\right)^2=32\lambda \frac{\mu^2\b-\omega^2}{\b^2}.
\ee

We see that the mass of the conjectured breather solution increases with a small $\lambda$ from the mass of the undeformed solution.

%%%%%%%%%%%%%%%%%%

\section{Shockwave solutions}

In this section, we consider the perturbative solutions of the shockwave type, more precisely shockwaves depending on 
$s\equiv t^2-x^2=x^+x^-$ ($x^\pm =t\pm x$), such that $\phi=0$ for $s<0$ (outside the light-cone coming out of $(x,t)=(0,0)$), 
and nontrivial only inside the light-cone, i.e., for $s\geq 0$. This is the case considered by Heisenberg for the action (\ref{Heisaction}).

\subsection{Shockwaves of the Heisenberg model}

The first case is of the Heisenberg model, with a general potential $V$, as in (\ref{Heisenberg}). 
This has been considered in \cite{Nastase:2015ixa}, and we just review it here. 
On the ansatz $\phi=\phi(s)$, 
we have
\be
X=(\d_\mu\phi)^2=-4s\left(\frac{d\phi}{ds}\right)^2\l,
\ee
so the Lagrangian on the ansatz is 
\be
l^2{\cal L}=1-\sqrt{1-l^4 4s \left(\frac{d\phi}{ds}\right)^2+l^42V}.
\ee

We will see that on the solution near $s=0$, which is the only one we will consider, the potential is negligible, so we will drop it for 
the moment. Afterwards we will see that this was self-consistent, since it is irrelevant for the solution. 

The equation of motion of the Lagrangian on the ansatz is 
\bea
0&=&2l^4\frac{d}{ds}\left[\frac{1}{\sqrt{1-4l^4 s\left(\frac{d\phi}{ds}\right)^2}}s\left(\frac{d\phi}{ds}\right)^2\right]\cr
&=&\frac{2l^4\left(1-2l^4s\left(\frac{d\phi}{ds}\right)^2\right)}{\left(1-4l^4 s\left(\frac{d\phi}{ds}\right)^2\right)^{3/2}}
\frac{d}{ds}\left[s\left(\frac{d\phi}{ds}\right)^2\right].
\eea

We see that the equation of motion reduces, near $s=0$, to 
\be
\frac{d}{ds}\left[s\left(\frac{d\phi}{ds}\right)^2\right]=0\;,
\ee
solved by 
\be
\phi=l^{-2}\sqrt{s} +{\cal O}(s^{3/2}).\label{heissol}
\ee

Then, {\em a posteriori}, we can check that indeed, if $V(\phi=0)=0$, $V(\phi(s))\simeq 0$ near $s=0$, so it can be 
neglected in the Heisenberg Lagrangian on the ansatz, and the solution above is still valid for $V\neq 0$. 

Moreover, we note that $X=(\d_\mu\phi)^2=-4s\left(\frac{d\phi}{ds}\right)^2$ jumps from 0 at $s=0-$ to $-l^{-4}$ at $s=0+$, 
even though $\phi(s)$ is continuous. 

In \cite{Nastase:2015ixa}, several further generalizations of this model have been considered, for instance with a function of 
$\phi$ in front of the $(\d_\mu\phi)^2$ inside the square root. There, however, we have seen that we cannot truncate the square 
root to any finite order; if we do so, we have no solution (\ref{heissol}) anymore. This, together with the fact that a canonical 
scalar with a potential also results in no solution (\ref{heissol}) was described as a certain "uniqueness" of the Heisenberg 
Lagrangian. 

We will show in Appendix A that, in fact, we can also exchange the square root for a fractional power smaller than 1, 
as well as consider $[(\d_\mu\phi)^2]^{2k+1}$ inside the square root, and an overall power $1/(2k+1)$, and the nontrivial 
shockwave (\ref{heissol}) is still a solution. However, only the first case is physically interesting, since in the second 
we obtain a complex on-shell Lagrangian.

Next, however, we consider the case of our $T\bar T$ deformation. 

\subsection{Shockwaves of the $T\bar T$ deformed  model}

Consider then, like in the case of \cite{Nastase:2015ixa}, that the scalar field $\phi$ is only a function of 
$s=t^2-x^2=(t+x)(t-x)$, $\phi=\phi(s)$, and also $\phi=0$ for $s\leq 0$, which means a propagating 
shockwave solution (instead of the general
behaviour of both $x+t$ and $x-t$, now we have only the dependence on their product). Then, as in the previous subsection,
\be
4\d \phi \bar \d \phi=-4s \left(\frac{d\phi}{ds}\right)^2\;,
\ee
and for simplicity we use the definition from (\ref{Lagra}),
\be
\tilde V\equiv \frac{2\lambda V-1}{2\lambda(1-\lambda V)}.
\ee

Then, the Lagrangian on the ansatz is 
\be
-{\cal L}
=\tilde V(\phi(s))+\frac{\sqrt{1+8\lambda(1-\lambda V(\phi(s)))(-4s)\left(\frac{d\phi}{ds}\right)^2}}
{2\lambda(1-\lambda V(\phi(s)))}.
\ee

Its equation of motion is
\bea
&&\tilde V'(\phi(s))+\frac{\sqrt{1+8\lambda(1-\lambda V(\phi(s)))(-4s)\left(\frac{d\phi}{ds}\right)^2}}
{2\lambda(1-\lambda V(\phi(s)))^2}\lambda V'(\phi(s))\cr
&&+\frac{-1/2}{2\lambda(1-\lambda V)}\frac{-8\lambda^2 V'(\phi(s))(-4s)\left(\frac{d\phi}{ds}\right)^2}
{\sqrt{1+8\lambda(1-\lambda V(\phi(s)))(-4s)\left(\frac{d\phi}{ds}\right)^2}}\cr
&&+\frac{d}{ds}\left[\frac{+8\lambda (1-\lambda V)(-8s)\frac{d\phi}{ds}}{2\lambda(1-\lambda V)
\sqrt{1+8\lambda(1-\lambda V(\phi(s)))(-4s)\left(\frac{d\phi}{ds}\right)^2}}\right]=0.
\eea

We want to see if the same solution near $s=0$, $\phi\simeq l^{-2} \sqrt{s}$, is valid here
(the coefficient of $X=(\d_\mu\phi)^2$ inside the square root is defined to be $l^4$).
We note that then, $s(d\phi/ds)^2\sim {\cal O}(1)$, whereas, assuming that $V(\phi)$ has only 
positive powers of $\phi$ (and no linear one, since that would be a tadpole in QFT, and is any way 
not something we want), $V(\phi)$ and $V'(\phi)$ go to zero on the solution near $s=0$. 
Then, it follows (as we can easily check) that the first two lines in the equation of motion above are 
subleading with respect to the third one. Moreover, in the third line, we can put $V$ to zero
for the leading behaviour, which finally just leaves the equation of motion of the Heisenberg 
model, so indeed its solution near $s=0$, $\phi\simeq l^{-2}\sqrt{s}$ (here $l^4=2\lambda$), is 
also a solution near $s=0$ here. We could have made a simpler argument: since, as we saw, $V(\phi)$ 
on the solution near $s=0$ vanishes, we could put that directly in the $T\bar T$ deformed action 
(\ref{Lagra}), which directly gives the Heisenberg DBI action (at $m=0$, since near $s=0$, Heisenberg 
already noted that the solution is independent of $m$ being or not zero), hence its solution, too.

\section{Conclusions}

%%%%%%%%%%%%%%%%%%%%%%%%%%%%%%%%%%%%%%%%%%%%%%%%%%%%%%%%%%%%%%%%%%%%%%%%%%%%%%%%%%%%%%%%

In this paper we have found solutions of the $T\bar T$ deformations and the Heisenberg deformation of the canonical scalar 
with a potential $V$. 

We have first found that the 1+1 dimensional $T\bar T$ deformation of a canonical scalar has soliton solutions, as well as shockwave 
solutions, that could be used in the Heisenberg model. We have written explicitly the static soliton solutions of the $T\bar T$
deformation action, in the case of 
sine-Gordon and Higgs-type potential $V$, where we have seen that the solitons are deformations of the solitons in canonical case.
We have also argued for the existence of breather solutions in the sine-Gordon case, also as deformations of the breather
solutions in the canonical case. In the generic case of the $T\bar T$ deformation solitons, we found the remarkable property that 
the mass of the solitons is undeformed, as is the solution near $\pm \infty$. 

In the case of shockwave solutions, we have shown that the Heisenberg perturbative shockwave 
solution $\phi(s)\simeq l^{-2}\sqrt{s}$ near $s=0$, with $\phi(s<0)=0$, is still valid for the $T\bar T$ case, as well as in other cases. 
Also a generic solution $\phi(x)=K|x-x_0|$ can be (infinitely) boosted to a solution $\phi(x^-)=\tilde K |x^-|$. 

The application of these solutions to the understanding of the Heisenberg model will be described in a separate publication
\cite{Nastase:2020}. 

There are several open questions that could be addressed  as a continuation of this research work. There include
\begin{itemize}
\item
The fact that the soliton mass of any deformed theory is the same as that of the undeformed theory  deserves 
further investigation. The question is whether there is some physical reason for that property and 
whether it has implications about other properties.
\item
This note includes a conjectured solution for the breather mode. Further work is needed to verify or falsify this conjecture.
\item
We have seen that there  is a shock wave solution for the $T\bar T$ system  has the same 
behavior as the solution of Heisenberg's model close to the origin. It will be interesting to 
further explore possible differences between the solutions in the other parts of space-time.
\item
A very natural generalization of this investigation is about  solutions of scalar field theories in higher space-time dimensions.
\end{itemize}

\section*{Note added}

After the paper was first posted on the arXiv, we became aware of the papers \cite{Conti:2018jho,Conti:2018tca}, which
have some overlap with the current paper, as they also discuss solitons, breathers and shockwaves in the context of $T\bar T$
deformations of the sine-Gordon model. According to the anonymous referee's suggestion, we have drawn the Figs. 2 and 7,8,9, 
in order to facilitate comparison with the breather solution in Fig.4 of \cite{Conti:2018tca} (in their case, as in ours, there is no 
analytical solution possible for $\phi(x,t)$, only a numerical one; in our case the implicit analytical solution in (\ref{conjecture}) cannot
be inverted).

\section*{Acknowledgements}

%%%%%%%%%%%%%%%%%%%%%%%%%%%%%%%%%%%%%%%%%%%%%%%%%%%%%%%%%%%%%%%%%%%%%%%%%%%%%%%%%%%%%%%%

We thank Aki Hashimoto 
for useful discussions. The work of HN is supported in part by CNPq grant 301491/2019-4 and FAPESP grants 2019/21281-4 
and 2019/13231-7. HN would also like to thank the ICTP-SAIFR for their support through FAPESP grant 2016/01343-7. The  work 
was of JS supported in part by a center of excellence supported by the Israel Science Foundation (grant number 2289/18).

\appendix

\section{Shockwave solutions in generalizations of the DBI and Heisenberg Lagrangians, with different powers}

In this Appendix we find that two possible modifications of the DBI and Heisenberg Lagrangians preserve the perturbative 
shockwave solution (\ref{heissol}). As in the main text, since we assume $V(\phi=0)=0$, and because (\ref{heissol}) is 
approximately zero near $s=0$, we neglect the potential, so we treat the modification of the Heisenberg Lagrangian as 
a modification of just the DBI Lagrangian. 

The first modification we analyze is 
\be
-{\cal L}l^4=\left[\sqrt{1+\left[l^4(\d_\mu\phi)^2\right]^{2k+1}}-1\right]^{\frac{1}{2k+1}}.
\ee

We note that the overall power was chosen so that, for $l\rightarrow 0$, we obtain the canonical Lagrangian, 
${\cal L}=-(\d_\mu\phi)^2/2$. The power of $X$ inside the Lagrangian was chosen such that $(-1)^{2k+1}=-1$.

As before, for $\phi=\phi(s)$, $X\equiv (\d_\mu\phi)^2=-4s\left(\frac{d\phi}{ds}\right)^2$, so the Lagrangian on the ansatz is 
\be
-{\cal L}=\left[\sqrt{1+\left[-l^44s\left(\frac{d\phi}{ds}\right)^2\right]^{2k+1}}-1\right]^{\frac{1}{2k+1}}.
\ee

The equation of motion is 
\bea
&&\frac{d}{ds}\left\{
\left[\sqrt{1+\left[-l^44s\left(\frac{d\phi}{ds}\right)^2\right]^{2k+1}}-1\right]^{\frac{1}{2k+1}-1}
\left\{\frac{2k+1}{\sqrt{1+\left[-l^44s\left(\frac{d\phi}{ds}\right)^2\right]^{2k+1}}}\times\right.\right.\cr
&&\left.\left.\times \left[-4s\left(\frac{d\phi}{ds}\right)^2\right]^{2k}\left(-8s\frac{d\phi}{ds}\right)\right\}\right\}=0.\label{ttdeomp}
\eea

Next, we want to check whether the solution (\ref{heissol}) near $s=0$ is still valid. As a first step, we note that, due to the 
properly chosen power inside the square root, the square root still vanishes for $\phi(s)=l^{-2}\sqrt{s}$. That means that, 
from among the terms in the equation of motion (\ref{ttdeomp}), the dominant one will be the one where the external $\frac{d}{ds}$
acts on the square root in the denominator, which means the equation of motion
\be
\frac{d}{ds}\left[s\left(\frac{d\phi}{ds}\right)^2\right]=0\;,\label{effeom}
\ee
indeed solved by (\ref{heissol}).

The only problem with this Lagrangian and its solution is that, on the solution, the Lagrangian is complex, since the square 
root vanishes, so
\be
{\cal L}_{\rm on-shell}=-(-1)^{\frac{1}{2k+1}}=-e^{\frac{\pi i}{2k+1}}\in \mathbb{C}.
\ee

The second modification is more physical, 
\be
-2\frac{p}{q}l^4{\cal L}'=\left[1+l^4(\d_\mu\phi)^2\right]^{\frac{p}{q}}-1\;,
\ee
where the overall constant in the Lagrangian was chosen such that the first term in the expansion in $l^4$ is the canonical 
kinetic term $-\frac{1}{2}(\d_\mu\phi)^2$. In order for the Lagrangian to appear to a negative power in the equation of motion, 
we choose $p/q<1$, i.e., $p<q$.

On the ansatz $\phi=\phi(s)$, the Lagrangian is 
\be
-2\frac{p}{q}l^4{\cal L}'=\left[1-l^44s\left(\frac{d\phi}{ds}\right)^2\right]^{\frac{p}{q}}-1\;,
\ee
and the equation of motion is 
\be
\frac{d}{ds}\left\{\left[1-l^44s\left(\frac{d\phi}{ds}\right)^2\right]^{\frac{p}{q}-1}\left(-8s\frac{d\phi}{ds}\right)\right\}=0.
\ee

As for the first Lagrangian, the square root vanishes on the solution (\ref{heissol}), and since in the 
equation of motion above it appears to the negative power $p/q-1$, the leading term in the equation on the 
solution is the one where the overall $d/ds$ acts on the square root, namely (\ref{effeom}), which does indeed have (\ref{heissol})
as a solution. 

Thus this Lagrangian still has the same perturbative shockwave solution, and this time the on-shell Lagrangian is actually 
real (and positive).

%%%%%%%%%%%%%%%%%%%%%
%\section{ On the search of a  soliton solution for the $T\bar T$ scalar field}
%\section{The derivation of the Hamiltonian density for the $T\bar T$ deformed action}

%The Hamiltonian density as written in (\ref{Lagrang})
%\be
%{\cal H}= -{\cal L}= \tilde V + \frac{\sqrt{1+2\bar \lambda \d_\mu \phi \d^\mu \phi}}{2\bar \lambda}\;,\label{Lagrangg}
%\ee
%From the solution of the equation of motion for $C=0$
%\be\label{solTbT}
%\frac{1}{2\lambda(1-\lambda V)\sqrt{1+2\lambda (1-\lambda V)\phi'^2}}+\tilde V=C\;,
%\ee
%We find that
%\be
%\sqrt{1+2\lambda (1-\lambda V)\phi'^2}= -\frac{1}{2\bar\lambda \tilde V}
%\ee 
%Therefore 
%\be
%{\cal H}= \tilde V - \frac{1}{4(\bar\lambda)^2 \tilde V} =\tilde V\left (  1-  \frac{1}{4(\bar\lambda \tilde V)^2}\right )
%\ee
%since $4 (\bar \lambda \tilde V)^2= (1-2\lambda V)^2$ we get
%\be
%{\cal H}= \tilde V\left (  1-\frac{1}{(1-2\lambda V)^2}\right ) =-\tilde V\left (\frac{4\lambda V( 1-\lambda V)}{(1-2\lambda V)^2}\right )\ee
%Substituting the expression for $\tilde V$ we get
%\be
%{\cal H}=\frac{(1-2 \lambda V)}{2\lambda( 1-\lambda V)} \frac{4\lambda V(1-\lambda V)}{(1-2\lambda V)^2}= \frac{ 2V}{1-2\lambda V}
%\ee
%%%%%%%%%%%%%%%%%%%%%%%%%%%%%%%%%%%%%%%%%%%%%%%%%%%%%%%%%%%%%%%%%%%%%%%%%%%%%%%%%%%%%%%%
\bibliography{DBITTbar}
\bibliographystyle{utphys}
%%%%%%%%%%%%%%%%%%%%%%%%%%%%%%%%%%%%%%%%%%%%%%%%%%%%%%%%%%%%%%%%%%%%%%%%%%%%%%%%%%%%%%%%

\end{document}